\documentclass[aps,twocolumn,groupedaddress,showpacs]{revtex4}
\usepackage{graphics}
\begin{document}
\newcommand{\beq}{\begin{equation}}
\newcommand{\eeq}{\end{equation}}
\bibliographystyle{apsrev}

\title{Lagrangian in quantum mechanics is a connection one-form}

\author{Pankaj Sharan}
\email{pankaj@jamia.net}
       
\author{Pravabati Chingangbam}
\email{prava@jamia.net}
\affiliation{Department of Physics, Jamia Millia Islamia, New Delhi-110025,
INDIA.}


\newcommand{\ra}{\rightarrow}
\newcommand{\al}{\alpha}
\newcommand{\be}{\begin{eqnarray*}}
\newcommand{\ee}{\end{eqnarray*}}
\newcommand{\om}{\omega}
\newcommand{\Om}{\Omega}
\newcommand{\Z}{\langle}
\newcommand{\R}{\rangle}
\newcommand{\A}{\int_{\tau_1}^{\tau_2} A(\tau)\, d\tau}
\newcommand{\rr}{{\bf r}}
\newcommand{\vv}{{\bf v}}
\newcommand{\RR}{\mathcal R}
\newcommand{\HH}{\mathcal H}
\newcommand{\PP}{\mathcal P}
\newcommand{\UU}{\mathcal U}
\newcommand{\CC}{\mathcal C}
\newcommand{\sss}{\scriptscriptstyle}

\begin{abstract}
We recast Dirac's Lagrangian in quantum mechanics in 
the language of vector bundles and show that the action
is an operator-valued connection one-form. Phases associated with 
change of frames of reference are seen to be total differentials in the 
transformation of the action. The relativistic 
case is discussed and we  show that it gives the correct phase in the non-relativistic 
limit for uniform acceleration.
\end{abstract}


\maketitle

\section{Introduction} 
As expressed by Landau and Lifshitz \cite{landau} `Quantum theory 
occupies a very unusual place among physical theories: it contains 
classical mechanics as a limiting case, yet at the same time it 
requires this limiting case for its own formulation'.  
There can be only one correct theory in nature and that, from all 
indications, must be a quantum theory. 
Quantum thery begins with a classical Lagrangian, action  or 
Hamiltonian formulation. But it is the classical formulation that has to be 
explained in terms of quantum theory. The 
difficulties faced by the process of quantization, that is, attempting 
to build a quantum theory from classical theory compel us to look for 
quantum origin of fundamental quantities like the Lagrangian or the 
Hamiltonian structure. This was the attempt made by 
Dirac in a remarkable paper ``The Lagrangian in quantum mechanics" 
\cite{dirac}. In this paper Dirac proposed what should be regarded as 
the Lagrangian 
(and therefore action) in quantum mechanics. As is well known this paper 
led to Feynman's formulation of the path integral 
\cite{feynman}, as well as to the Schwinger action principle 
\cite{schwinger}. If one reads between the lines in Dirac's paper the picture of 
a vector bundle 
with base space, fibre and connection can be glimpsed there. 
The aim of this paper is to formulate this and identify the precise mathematical 
nature of the Lagrangian in quantum mechanics.

There has been a recent trend to view quantum 
evolution as a kind of parallel transport especially since the discovery of the 
geometric phase by Berry \cite{berry}. It was actually Asorey et al 
\cite{asorey} who first floated the idea of time evolution of a quantum system 
as a parallel transport, though their transport was limited to one dimensional 
manifold of time, without any of the interesting geometric consequences. 
This viewpoint becomes natural when we set up Dirac's argument in 
differential geometric language. 

The paper is organized as follows: in section 2 we present Dirac's argument 
in the language of vector bundles. In 
section 3 we define the quantum mechanical action and demonstrate the 
usefulness of the formalism when applied to arbitrary frames. 
In section 4 we extend our definition to the relativistic 
case and show that it gives the correct phase in the non-relativistic 
limit for uniform acceleration given.  
The minimum differential geometry needed for our construction is given 
in the Appendix.

\section{Dirac Lagrangian as a connection}

In \cite{dirac} Dirac took up the question  `what corresponds in the quantum 
theory to the Lagrangian method of the classical theory?' and he showed 
that the quantity $\langle \xi'',t+dt|\xi',t\rangle$ is what corresponds to 
$\exp(iLdt/\hbar)$, where $L$ is the Lagrangian. We proceed to reformulate 
his argument in a differential geometric language. 

Let $\{\xi(t)\}$ be a complete set of commuting observables in the Heisenberg
picture, giving a {\it moving representation} $\Z \xi',t| $ in terms of
eigenvectors of $\xi(t)$. The canonical momenta $P(t)$ and the Hamiltonian
$H$ act on these states as
\beq
 -i{\partial\over{\partial \xi'}}\Z \xi',t| = \Z \xi',t|\,P(t) 
\eeq
\beq
i{\partial\over{\partial t}}\Z \xi',t| = \Z \xi',t|\,H   
\eeq	
where $P(t)=e^{iHt}P(0)e^{-iHt}$.
We assume the set of eigenvalues $\xi'$ of $\xi$ forms a manifold $M$ and call
$ B=M\times R $ (where $R$ represents time) as {\it spacetime}.

The entire dynamics of the system may then be determined from an assignment 
of basis vectors
$\Z \xi',t| $ to each point $ (\xi',t) $ of $B$. One can think of a 
(generalized) Hilbert space provided at each point $(\xi',t)$.

Let $ c:\tau\ra (\xi'(\tau),t(\tau)) $ be a smooth curve in $B$. It
follows from Eqs.(1) and (2) that rate of change of the basic vector 
$\Z \xi',t|$
along the curve is
\beq
{d\over{d\tau}}\Z \xi'(\tau),t(\tau)|
      =i \Z \xi'(\tau),t(\tau)|\,\Big({{d\xi'}\over{d\tau}}P(t)
            -{{dt}\over{d\tau}}H \Big) .    
\eeq					
Denoting 
\beq
\omega^{\xi}\equiv -i\Big(P(t)d\xi'-Hdt\Big),       
\eeq
where the superscript $\xi$ indicates that the operator 
$\omega^{\xi}$ is determined by the choice of the 
complete set of commuting observables, we have
\beq
d\Z \xi',t|= - \Z \xi',t|\,\omega^{\xi}.   
\eeq
This equation can be thought of as a parallel transport, $D\Z \xi',t|= 0$, with
\beq
D=d+\omega^{\xi}  \eeq
provided we can show that $\omega^{\xi}$ transforms like a 
connection. 

For a finite change we have 
\beq
\langle \xi_2,t_2|=P\Big(\langle \xi_1,t_1|\exp\Big[-\int_{t_1}^{t_2}\omega^{\xi}
\Big]\Big), 
\eeq
where $P$ stands for path ordering along $\tau$. In this picture 
$\omega^{\xi} $ 
 is an operator-valued one-form which acting on a tangent to a curve
determines an operator.  It permits the comparison of vectors 
belonging to different fibres \cite{pank}.

In order to see the transformation property of $\omega^{\xi} $, let 
$\{ \eta'(t')\}$ be another complete set of commuting observables. Then 
the action associated with it is $\omega^{\eta}$ given by 
\beq d\Z \eta',t'| = -\Z \eta',t'|\omega^{\eta}. \eeq
By inserting a complete set 
$\int |\xi',t\rangle d\xi'\langle \xi',t| =1$ we also have
\begin{eqnarray*}
d\Z \eta',t'| &=& d\Big(\int \Z \eta',t'|\xi',t\R d\xi'\Z \xi',t| \Big)\\
{} &=&\int d\xi'\big( d\Z \eta',t'|\xi',t\R \big) \Z \xi',t|\\
{} &{}&  + \int d\xi' \Z \eta',t'|\xi',t\R d\Z \xi',t|\\
{} &=&\int d\xi'\big( d\Z \eta',t'|\xi',t\R \big) \Z \xi',t|\\
{} &{}&  - \int d\xi' \Z \eta',t'|\xi',t\R \Z \xi',t|\omega^{\xi}.
\end{eqnarray*}

Taking inner product with $|\eta'',t\R$ we get
\begin{eqnarray*}
-\Z \eta',t'|\omega^{\eta}|\eta'',t'\R 
   &=& \int d\xi' d\Z \eta',t'|\xi',t\R \Z \xi',t|\eta'',t'\R\\
{} &{}&    - \int d\xi' \Z \eta',t'|\xi',t\R \Z \xi',t|
                  \omega^{\xi}|\eta'',t'\R.
\end{eqnarray*}
Calling the unitary matrix $U_{\eta'\xi'}\equiv \Z\eta',t'|\xi',t\R$ 
we have
\beq
\omega^{\eta}_{\eta'\eta''} 
      = U_{\eta'\xi'}\omega^{\xi}{\xi'\xi''}U^{-1}_{\xi''\eta''} 
                       - dU_{\eta'\xi'}U^{-1}_{\xi'\eta''} 
\eeq		       
with summation and integration understood over repeated indices. 
The above equation can be written in index free notation as 
\beq \omega^{\eta} = U\omega^{\xi} U^{-1} + UdU^{-1}. \eeq
This is the transformation rule for a connection. 




The curvature is given by (see Appendix) 
\begin{eqnarray*}
 \Omega &=& d\omega +\omega\wedge\omega\\
{}      &=& -i\Big( {{\partial P(t)}\over{\partial t}} 
            + {{\partial H}\over{\partial \xi'}} + i[P,H]\Big)
	    d\xi'\wedge dt.
\end{eqnarray*}
For $H=P^2/2m+V(\xi)$ and using the Heisenberg equation of motion for 
$P(t)$ we get
\beq
\Omega = -i {{\partial V}\over{\partial \xi'}}  d\xi'\wedge dt  
\eeq
which indicates that force is curvature.

Thus, by setting up a fibre bundle construction we have shown that the 
action in quantum mechanics must be a connection. It governs the 
parallel transport of Hilbert space vectors along arbitrary curves in 
the bundle which are lifted from the base manifold.

\section{Phase changes due to change of frames}

In this section we use the above picture and construct a Hilbert vector 
bundle with  
physical spacetime as the base manifold. Quantum mechanically this 
amounts to singling out the position operators as the preferred 
complete set of observables.  We then determine what  
$\omega$ should be in different frames of reference. It 
turns out that this formulation gives a natural explanation of the 
phases that must accompany changes of frames of reference. In the usual 
method the transformation properties of the Schr\"odinger equation is 
first guessed and then phases are fixed to satisfy the correct 
Schr\"odinger equation. This happens for Galilean transformation 
as well as for accelerating frames. 

Here, we show that whenever there is a 
change of frame of reference, treated 
as a change of coordinates on the base manifold,   
re-expression of the action connection in the coordinates 
of the new frame gives rise to total differentials 
or exact forms $d\phi(x,t)$. These exact forms become phase 
factors $\exp(i\phi(x,t))$ under parallel transport given by Eq. (7). 

Let us consider a one-particle system and let the position degrees of freedom 
constitute a complete set of commuting operators. Let us identify the 
eigenvalues of the position operators with physical space.  
Consider the vector bundle whose typical fibre is Hilbert space $\HH$, base 
manifold is spacetime and the structure group is the group of all unitary 
operators on $\HH$. Let 
$P_{\mu}=(H,-{\vec P})$ be the generators of spacetime translation operators 
which act on the Hilbert space at each point, where ${\vec P}=P^i=-P_i$ 
is the physical momentum.

We now propose that : {\it the operator-valued connection one-form 
$ \omega=iP_{\mu}dx^{\mu}$ be called the quantum mechanical action.}

In the following we investigate the transformation of $\omega$ 
under change of frames of reference. 

\subsection{Uniformly moving frames}

Uniformly moving 
frames can be accommodated by a change of coordinate from static to 
moving ones.
\begin{eqnarray}
{x'}^0  &=& x^0=t  \\
{x'}^i  &=& x^i-v^i t
\end{eqnarray}
The connection form $\omega$ when expressed in terms of $x'$ looks like 
\beq
i\omega = p_id{x'}^i -{{1}\over{2m}} \Big( p^2-2m v^i p^i \Big)dt'.  
\eeq
The momentum in the primed coordinates must be ${p'}^i=p^i -mv^i$.  
We next put $i\omega$ in a form which emphasizes the equivalence of the two 
coordinate systems, as follows
\be
i\omega &=& {p'}_id{x'}^i -{{{p'}^2}\over{2m}} 
             +d\Big( m v^i {x'}^i + {{1}\over{2}}mv^2t'\Big)  \\
{}      &=& i\omega' + d\Big[ m v^i {x'}^i + {{1}\over{2}}mv^2t'\Big]     
\ee	   
where $i\omega'={p'}_id{x'}^i -p'^2/2m$. 
Thus, the action connection as seen in the two frames differ by an exact 
differential. This exact differential shows up in the wavefunction as 
a phase 
\beq \psi(x,t) =\exp\Big[ i\Big(  
                   m v^i{x'}^i+{{1}\over{2}}mv^2t' \Big)\Big]\psi'(x',t').\eeq

Therefore, parallel transport in the $(x',t')$ 
coordinates takes place with an additional factor.
This phase is well known from the representation theory of the  
Galilean group \cite{azcarraga}. It is precisely the phase that must be multiplied to 
the wavefunction in order to make the Schr\"odinger equation transform 
covariantly under Galilean transformation. It is worth emphasizing that 
no such assumption regarding how the Schr\"odinger equation must transform 
is required.

\subsection{Uniformly accelerated frames}

We can deal with accelerating frames also by simply treating them as 
change of coordinates and imposing energy-momentum relation in the new frame. 
It is sufficient to consider acceleration in the $x'$ direction. 
\begin{eqnarray}
x'  &=& x-{{1}\over{2}} gt^2,  \\
t' &=& t.
\end{eqnarray}

Proceeding in a similar manner as in the last section this gives
\beq
i\omega = pdx'+pgt'dt'-{{p^2}\over{2m}} dt'.
\eeq
The momentum in the primed coordinates must be ${p'}^i=p^i -mgt'$. 
Completing the square for the coefficient of $dt'$, which amounts to 
imposing the mass-shell condition in the accelerated frame we get
\begin{eqnarray*}
i\omega &=& {p'}_id{x'}^i -\Big( {{{p'}^2}\over{2m}} +mgx'\Big)dt'\\
{} &{}& +d\Big[ m g( t'x' + {{1}\over{6}}g{t'}^3)\Big]. 
\end{eqnarray*}
\beq
\eeq

Thus by simply re-expressing the action in the new frame and imposing 
the  energy-momentum relation we have obtained the pseudo-gravitational 
force field given by the potential $mgx'$ and the phase 
$m g( t'x' + {{1}\over{6}}g{t'}^3)$. It follows that the 
equivalence principle must hold in quantum mechanics as a natural 
consequence of the transformation property of the action. 

It must be pointed out that in the 
conventional approach \cite{eliezer} one demands that the transformation 
of the Schr\"odinger equation have an intuitive linear 
``gravitational potential term" and then it is found that the 
wavefunction must pick up the time dependent phase given above. 
The extra linear potential has been experimentally verified \cite{cow}.

It is interesting to see the group property of linear acceleration in 
the same direction. Let
\begin{eqnarray}
x'' &=& x'^1-{{1}\over{2}}g't'^2 \\
t'' &=& t.
\end{eqnarray}
The exact forms add giving 
$$ mg'(x''t''+{{1}\over{6}}g'{t''}^3)+mg(x't'+{{1}\over{6}}g{t'}^3)$$
\begin{eqnarray*}
{}   &=& m(g+g')x''t'' \\
{} &{}&   + {{1}\over{6}}mt''^3(g'^2+g^2+3gg'). 
\end{eqnarray*}   
\beq
\eeq
From the time component of the connection we get  
\beq
-mgx'dt'-mg'x''dt''=-m(g+g')x''dt''- {{1}\over{6}}mgg't''^3.
\eeq
which gives the combined transformation with acceleration $g+g'$. 

\subsection{Uniformly rotating frame}

In this case  we restrict ourselves to two space 
dimensions for simplicity. 
Let the coordinates of the rotating frame be given by
\begin{eqnarray*}
x' &=& x\cos {\tilde\omega} t + y \sin {\tilde\omega} t\\
y' &=& -x\sin {\tilde\omega} t + y \cos {\tilde\omega} t\\
t'   &=& t,
\end{eqnarray*}
where $\tilde\omega$ being the angular velocity. 
Then 
\begin{eqnarray*}
dx &=& \cos {\tilde\omega} t'dx' - \sin {\tilde\omega} t' dy' 
-{\tilde\omega} y dt'\\
dy &=& \sin {\tilde\omega} t'dx' + \cos {\tilde\omega} t' dy' 
          + {\tilde\omega} x dt'.
\end{eqnarray*}
The action connection in the new coordinates is 
\beq
i\omega = {\vec p}\,'.d\vec x\,' - \Big({{\vec p\,'^2}\over{2m}} 
           - {\tilde\omega} J\Big) dt'
\eeq	   
where the momentum in the rotated frame is 
$p'= Rp$, $R$ being the rotation matrix and $J=(x'p'^2-y'p'^1)$. 
The terms in $J$ can be combined with the Hamiltonian 
${\vec p\,'^2}/2m$ to recover Coriolis and centrifugal force terms
\be i\omega&=& {\vec p}\,'.d\vec x\,' \\
{}         &{}&- \Big({{(p\,'^1
                 +m{\tilde\omega} x')^2}\over{2m}}
              + {{(p\,'^2-m{\tilde\omega} y')^2}\over{2m}}
           -{{1}\over{2}}m{\tilde\omega}^2 {\vec x\,'}^2\Big) dt'. \ee
The Coriolis force does no work as it is perpendicular to velocity. 
So it does not appear as a potential 
term, rather, it appears as a connection or a vector potential in momentum 
in the Hamiltonian.

\section{Relativistic uniformly accelerating frame and its 
non-relativistic limit}

Let the action connection in relativistic quantum mechanics be defined 
by the same formula as in the non-relativistic case:  
\beq
-i\omega=P_{\mu}dx^{\mu}
\eeq
where $P_{\mu}$'s are the translation operators of Minkowski spacetime. 
We are not requiring that a complete 
set of commuting observables be provided as the basic ingredient for 
defining action. In fact we can no longer make such a demand because in 
a relativistic setup, the  
position operators are not defined \cite{ali}. Our basic assumption is 
that Minkowski spacetime be given and we construct the Hilbert vector bundle 
with it is the base manifold. We transform to a uniformly accelerating 
frame. On taking the 
non-relativistic limit of the action we expect that we should obtain 
Eq. (17).  We show that this is indeed  so. 

Consider Hilbert space consisting of momentum wave functions $\psi(p)$ 
of a spinless particle of mass $m$ with the inner product
\beq
(\psi,\phi)=\int {{d^3p}\over{2p_0}}\psi(p)^*\phi(p).
\eeq
We explicitly introduce the velocity of light $c$ on the formulas for 
convenience of taking the non-relativistic limit.
\begin{eqnarray*}
x^{\mu} &=& (ct,x^1,x^2,x^3)\\
p^{\mu} &=& (E,p^1c,p^2c,p^3c).
\end{eqnarray*}

Let us consider an observer which is uniformly accelerating along the 
$x$-axis. This means that the acceleration in the `co-moving' frame is 
a fixed number $g$. It can be shown \cite{narli} that such an observer 
follows a trajectory 
\beq
x = A\cosh \tau,\quad ct = A\sinh \tau. 
\eeq
where $A$ is related to the acceleration $g$ by $g=c^2/A$ and $\tau/g $ 
is the proper time in the `co-moving' frame. In order to treat change 
of frames of reference as a change of coordinates we must consider a 
continuum of observers with $\tau$ and $A$ as variables. 

\begin{figure}
\resizebox{!}{2in}{\includegraphics{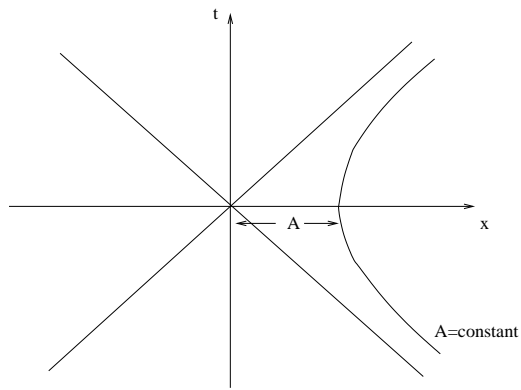}}
\caption{ Minkowski space and transformation to uniformly accelerated 
frame.}
\end{figure}

Thus we obtain a family of trajectories, one for each value of the 
acceleration $g$ or the relativistic accelerated space coordinate $A$ and 
parametrized by $\tau$. At $\tau=0$ we have $x=A$. 
At low velocities  
the non-relativistic limit for the trajectory is given by
\beq
x'=x-{{1}\over{2}}gt^2 .  
\eeq
This corresponds to large values of $A$.

In these coordinates $\omega$ is expressed as 
\beq
-i\omega =  P_{\tau}d\tau + P_AdA  
\eeq
where 
\begin{eqnarray}
P_{\tau} &=& A(P_0\cosh \tau + P_1\sinh\tau)\\
P_A &=& P_0\sinh \tau + P_1\cosh\tau.
\end{eqnarray}

In order to take non-relativistic limit we choose a suitably large $A$ 
such that
\beq
A=A_0+x',\ \ A_0={{c^2}\over{g_{\sss 0}}}, 
\eeq
where $A_0\gg x'$ and $g_{\sss 0}$ is the constant nonrelativistic 
acceleration. In the following we use $g$ for $g_{\sss 0}$. 
Then $(x',t)$ are non-relativistic uniformly accelerating 
coordinates and $\omega $ must be expressed in terms of these 
coordinates in the non-relativistic limit. We must make the 
following approximations
$$ P_0=mc^2 + {{p^2}\over{2m}}, \ \ P_1= -p  $$
$$ {{p}\over{mc}} \ll 1,\ \ {{gt}\over{c}} \ll 1 \ \
{\rm and} \ \ {{gx'}\over{c^2}} \ll 1.  $$
\beq \eeq

To get non-relativistic limit we must keep terms upto zeroth order only  
in the approximations. Since the terms $mc^2$  and $A$ are each of 
order $-1$ in degree of smallness we must at least keep terms upto 
second order before we multiply out all factors and throw away terms 
higher than zeroth order. 
Thus with these approximations we get 
\begin{eqnarray*}
P_{\tau}d\tau &=& \Big( mc^2 + {{p^2}\over{2m}} -pgt \Big)cdt\\
{}            &=&  +mgx'cdt-mgx'cdt-mgtcdx'\\
{}            &=& \Big( mc^2 + {{p^2}\over{2m}} -pgt +mgx'\Big)cdt
		    -cd(mgx't). 
\end{eqnarray*}
\beq \eeq

Upon completing the square to get $(p-mgt)^2/2m$ we get
\be
P_{\tau}d\tau &=& \Big( mc^2 + {{(p-mgt)^2}\over{2m}} +mgx'\Big)cdt\\
{}&{}&  -cd\Big[mg(x't + {{1}\over{6}}gt^3)\Big].   
\ee
\beq \eeq

The other term in the action is
\beq
P_AdA = -\Big(p-mgt \Big) cdx'. 
\eeq
Thus the non-relativistic action connection is obtained to be
\be
 -i\omega &=& \Big( mc^2 + {{(p-mgt)^2}\over{2m}} +mgx'\Big)cdt\\
{} &{}&        	  -\Big(p-mgt \Big) cdx' 
		  -cd\Big[mg(x't + {{1}\over{6}}gt^3)\Big]. 
\ee
\beq \eeq
Comparison with Eq. (17) immediately tells that this is the correct 
expression for the action. 
 


\appendix

\section{Geometric setting}

The material in this appendix is only for fixing notation. The geometry 
is explained in well known books, see for example Chern et al 
\cite{chern}.

\subsection{The bundle and connection}

Consider a vector bundle $E$ with base manifold $M$, Hilbert space $\mathcal H$ 
as fibre 
and the group $\mathcal U$ of all unitary transformations on $\mathcal H$ 
as the structure group.

Let $\phi_n(x)$  be a set of smooth sections such that it forms an 
orthonormal basis in the fibre at $x$.
\beq
\big(\phi_n(x),\phi_m(x) \big)=\delta_{nm} . 
\eeq
Any arbitrary section $\psi(x)$ can then be written as
\beq
\psi(x)= c_n(x)\phi_n(x)  
\eeq
where $c_n(x)$ are the complex coefficients of expansion and we 
use Einstein summation convention.
Let $\Gamma$ be the vector space of all sections. They can be added pointwise.
\beq
\big( \psi_1+\psi_2 \big)(x)=\psi_1(x)+\psi_2(x) 
\eeq
and multiplied with smooth functions 
\beq
(c\psi)(x)=c(x)\psi(x) .
\eeq
Let $\Lambda\otimes\Gamma$ be the tensor product of the space $\Lambda$ 
of all one-forms on the base $M$ and $\Gamma$. A connection on this 
bundle is a mapping $D:\Gamma \ra \Lambda\otimes\Gamma $ such that
\beq
D(\psi_1+\psi_2) = D\psi_1+D\psi_2   
\eeq
and
\beq
D(c\psi)         = cD\psi +dc\otimes\psi  .
\eeq
As $\phi_n(x)$ is a basis in $\Gamma$ we can express $D(\phi_n)$ in 
terms of the basis $dx^\mu \otimes \phi_m$ in $\Lambda\otimes\Gamma$ as
\beq
(D\phi_n)(x) = \phi_m(x) \Gamma_{\mu m n} dx^\mu
\eeq
where coefficients $ \Gamma_{\mu mn}(x)$ are the 
Christoffel symbols with respect to the basis  $dx^\mu\otimes\phi_m$. 
We write this equation as 
\beq
(D(\phi_n)(x) = \phi_m\om^{\phi}{}_{mn}(x), 
\eeq
where the complex matrix $\omega_{mn}$ can be obtained by taking inner 
product with  $\phi_m$ in Eq. (6).
\beq
\om^{\phi}_{mn} = \big( \phi_m,D\phi_n\big) .
\eeq
This matrix of one-forms is called the {\em connection 
matrix}. 

For $\psi(x)=c_m(x)\phi_m(x)$ we have
\beq
\big(\phi_n,D\psi\big)=dc_n + \Gamma_{\mu nm}c_mdx^{\mu} 
                              = dc_n + \omega_{nm}c_m  .
\eeq			      

We require $D$ to satisfy the Leibniz rule 
\beq
D(\phi,\psi) = (D\phi,\psi) + (\phi,D\psi) = d(\phi,\psi), 
\eeq
which when applied to $\delta_{mn} = (\phi_m,\phi_n)$ shows
that $\om^{\phi}$ is an anti-hermitian matrix.

\noindent Under a change of basis 
\beq
\chi_n(x) = U(x) \phi_n(x)  
\eeq
we have
\be
\chi_n(x)  &=& \phi_m(x)\big( \phi_m(x), U(x) \phi_n(x)\big)  \\
           &=& \phi_m(x)U_{mn}(x).  
\ee	   
\beq \eeq
\noindent Omitting the base point $x$ for simplicity of notation
\be
D\chi_n  &=& D\big(\phi_s U_{sn}\big) \\
              &=& \phi_r \om^{\phi}_{rs}U_{sn} + \phi_s dU_{sn} \\
              &=& \chi_m\om^{\chi}_{mn} = \phi_r U_{rm} 
	      \om^{\chi}_{mn}                  	       	     
\ee
\beq \eeq
Or
\beq
\om^{\chi}_{mn} = U^{-1}_{mr} \om^{\phi}_{rs}U_{sn} 
                       + U^{-1}_{mr}dU_{rn} .    \eeq
Omitting matrix indices, we have		       
\beq \om^{\chi} = U^{-1} \om^{\phi}U + U^{-1}dU .\eeq
The curvature two-form for the connection is given by
\beq \Om^{\phi} = d\om^{\phi} + \om^{\phi} \wedge \om^{\phi},  \eeq
which transforms as 
\beq \Om^\chi = U^{-1}\Om^{\phi}U . \eeq

\subsection{Parallel transport}

Let 
\beq  x^{\mu} = c^{\mu}(\tau)   \eeq
 be a curve in the base manifold. Let $\psi(x)$
be a (Hilbert) vector field  parallel transported along 
$ c^{\mu}(\tau)$. This means
\beq <\dot{c},D\psi> =0\quad {\rm{for\ all\ \tau}}.  \eeq

Then
\be
<\dot{c},D\psi> &=&< {{dc^{\mu}}\over{d\tau}} 
                      {{\partial}\over{\partial x^{\mu}}}
		      da_m \phi_m + i \phi_m P_{\mu mn}a_n >  \\
	{}	 &=&  {{dc^{\mu}}\over{d\tau}} \Big(
		     {{\partial a_m}\over{\partial x^{\mu}}} \phi_m 
		     + i \phi_m P_{\mu mn}a_n    \Big)\\
{}	         &=& 0	     
\ee
\beq \eeq
which gives
\beq i \partial_{\mu} a_m = P_{\mu mn} a_n. \eeq
Or,
\beq a_m(y) = P\Big(\exp\Big(-i\int_x^y P_{\mu} dx^{\mu}\Big)_{mn} 
                      a_n(x) \Big)    \eeq
where $P$ stands for path ordering.

When the constant basis in the common Hilbert space is understood we 
need not specify vectors and operators as matrices and use operator 
notation. For example the equation above can be written as
\beq \psi(y) = P\Big[\exp\Big(-i\int_x^y P_{\mu} dx^{\mu}\Big) 
                      \psi(x) \Big]  . \eeq

\subsection{Curvature}
If the connection is written as
$$ \omega= iP_{\mu} dx^{\mu}   $$
then the  curvature is
\beq \Omega =  {{i}\over{2}} F_{\nu\mu}dx^{\nu}\wedge dx^{\mu} \eeq
where 
\beq F_{\mu\nu}= \partial_{\mu}P_{\nu} - 
            \partial_{\nu}P_{\mu} + i[P_{\mu},P_{\nu}]. \eeq

\end{document}